\documentclass[11pt, oneside]{amsart}   	% use "amsart" instead of "article" for AMSLaTeX format
\usepackage{geometry}                		% See geometry.pdf to learn the layout options. There are lots.
\usepackage{amssymb}
\usepackage{mathrsfs}
\begin{document}
%\maketitle 
\begin{center}
Surprisingly Long Length Scales for Semiclassical Loop Quantum Gravity and Their Physical Consequences
\\ 
\vspace{2ex}
P. G. N. de Vegvar \\ 
\vspace{2ex}
SWK Research\\
24412 McMurray Ridge Lane, Mount Vernon, Washington, USA\\
  \vspace{2ex}
Paul.deVegvar@post.harvard.edu\\ 
\vspace{2ex}
PACS: 04.60.Pp, 98.80.Cq 
\end{center}
%\begin{abstract}
When gauge field theory coherent states for loop quantum gravity (LQG) were introduced, an optimized semiclassical proper length emerged, corresponding to the edge length $\epsilon $ of a graph embedded in a given classical
geometry. Here $\epsilon$ is explored in more detail. $\epsilon $ at the Earth's surface is found to lie between 100 $\mu $m and 0.7 m.  The implied
quantum fluctuating space-time strain amplitude and noise spectrum are estimated to be $4\; 1/2$
orders smaller than the current experimental detectability. However, such a macroscopic $\epsilon $ makes regularization of
the semiclassical electromagnetic Hamiltonian problematic for photon wavelengths shorter than $\epsilon $.  The origin of a large  $\epsilon $ is traced to an edge-wise tensor product of independent edge-based coherent states for the whole graph state.  This provides physical grounds
for recently proposed collective coherent states, where $\epsilon $  acquires the interpretation of a sliding scale. A new proper distance $\xi$ emerges  as the characteristic length of semiclassical LQG. $\xi$ will affect the LQG photon vacuum dispersion relations, and is also accessible to current measurements of space-time strain. Matter interactions may also affect $\xi$.
\vspace{2.8in}
%\end{abstract}
\pagebreak

\section{Introduction}
%\subsection{}
Over the last two decades loop quantum gravity (LQG) has emerged as an important and mathematically rigorous approach to canonical quantum gravity.\cite{Rovelli}\cite{Th_Bk} While much of the effort is directed towards Planck scale physics, the semiclassical regime is not only attracting theoretical interest, but gravitational phenomena of low order in $\hbar $ are also promising places to search experimentally for novel physics and to test the predictions of LQG.\cite{GCS}\cite{Sahl_Thesis}\cite{Flux_Repn_Coh_St} One well known (and controversial) effort along these lines is the on-going test for vacuum dispersion of photons propagating over long distances.\cite{GP}\cite{ATU}\cite{Fermi_LAT}  Somewhat related to this, but less explored, is the notion of an optimized edge length
for LQG graphs embedded in a given classical geometry. This idea was first proposed by Sahlmann, Thiemann,  and Winkler (STW) \cite{ITP} when they applied the mathematics of coherent states on compact groups\cite{Hall} to LQG.  A theoretical prediction of a new characteristic length or energy scale such as this often presages potentially new and interesting physics. The electroweak scale is one instance of this from particle physics. Condensed matter abounds with examples, such as critical point correlation lengths, superconducting coherence and magnetic field penetration lengths, to mention just a few. In this article we present some consequences of treating the LQG graph edge length as a gravitational scale laying between the Planck scale $L_P$ and the classical curvature radius $L_C$. Although direct experimental probes of quantum fluctuating  space/space-time along graph edges are out of reach of current technology, a few simple estimates can remarkably offer guidance to ongoing theoretical developments in coherent state construction. In fact, one learns that such an edge-based gravitational mesoscopic regime can already be precluded and another length scale replaces the edge length in delineating that regime. There are consequences of this new length for the LQG photon vacuum dispersion relations as well as for current measurements of space-time strain. \\

Since the focus here is on presenting the results of applying existing theory, the background material is only briefly sketched, trusting the reader to be reasonably familiar with LQG.\cite{Rovelli}\cite{Th_Bk} Next we estimate the optimized embedded graph edge length at the Earth's surface using the same method as \cite{ITP}, and study the quantum fluctuations associated with it. Then we show the mere fact the optimum edge length has a surprisingly large value on Earth poses a problem for regulating the semiclassical matter sector. Finally we point out how to circumvent this problem confronting the construction of  the graph-based coherent states. We conclude by commenting on what the edge length $\epsilon$ and a new collective geometric correlation length $\xi$ mean physically and how $\xi$ may be experimentally probed. Finally, we speculate on how the matter sector could play a role in determining $\xi$. \\

\section{Brief Sketch of LQG}
Here we use the LQG canonical general relativity approach to gravitational phase space (without matter degrees of freedom).\cite{Rovelli}\cite{Th_Bk} This utilizes the intrinsic 3-metric
$q_{ab}$ and extrinsic curvature $K_{ab}$ of the 3-dimensional (compact, connected)  (initial) spatial slice $\Sigma$ of the ADM spacetime decomposition.\cite{MTW}
These metric based variables are replaced by real-valued 
connections $A^i_a$ and densitized dual triads $E^a_j$, which live on $\Sigma$ , and these are then integrated (smeared) over the edges $e$ of finite graphs and their dual elementary 2-surfaces $S_e$, respectively. The connection describes a frame bundle with reduced gauge group SO(3) or SU(2). The fundamental classical and quantum dynamical quantities are the group-valued holonomies $h_e$ of the connection along the graph's edges and the fluxes
\begin{equation}
 p^{e}_{j} = -\left(\frac{1}{a^2}\right) \mathrm{Tr}  \left[ \tau _j h_e (0,1/2) \left( \int_{S_e} h_{\rho _{e}(x)} \left( *E(x) \right) h_{\rho _{e} (x)}^{-1} \right) h_{e} ^{-1} (0,1/2) \right] \label{PDEFN} \\
\end{equation}
where $h_{e} (0, 1/2)$ is the holonomy from the starting point of the edge to its midpoint (where $S_e$ intersects edge $e$), and $\rho_{e} (x)$ is a path in $S_{e}$ from that intersection point to $x$.  Here $a$ is a quantity with dimensions of length, so both $p^{e}_{j}$ and $h_{e}$ are dimensionless, but $a$ is initially left arbitrary. $\{\tau_{j}\}$ is a set of generators for the Lie algebra. Under quantization in the connection representation, functions of the holonomies become operators which act on Hilbert spaces
of cylindrical functions of the edge holonomies by multiplication, and the fluxes $p^{e}_{j}$ become operators that act as invariant vector fields on cylindrical functions.  Starting with earlier work by Hall\cite{Hall}, Thiemann,  Sahlmann, and Winkler constructed coherent (semiclassical) states in the connection representation based on a single copy of the compact gauge group SU(2). These are kinematic coherent states, and do not generally satisfy the gauge invariance (Gau\ss), 3-diffeomorphism, or Hamiltonian constraints. Nevertheless, the kinematic states are expected to capture much of the physics.\cite{GCS}\cite{Bahr} However, no coherent state for LQG  that satisfies the Hamiltonian constraint has been reported so far in the literature. Briefly, the idea is to choose a classical initial 3-geometry $m$  of $\Sigma$ from classical phase space $\mathscr{M}$ and then to construct a state $\psi _{m}$ associated with $m$ (a pair of configurations  $A^i_a$ and $E^a_j$ on
 $\Sigma$), which is "semiclassical" for elementary classical observables $g_e$. Specifically that requires:  (a) expectation values coincide with classical values
 \begin{equation}
  \langle\psi_ {m}, \hat g_{e}\psi_{m}\rangle _{e}(m)\approx g_{e}(m)\label{SemiclA}
  \end{equation}
   to at least 0-th order in $\hbar$, and (b) fluctuations are small, meaning
\begin{equation}
(\Delta g_{e})_{\mathrm{quant}} (m) = \left[ \frac{\langle\psi_{m}, (\hat {g}_{e})^{2} \psi _{m}\rangle}{(g_{e}(m))^{2}} - 1 \right] \ll 1, \label{SemiclB}
\end{equation}
or choose $g_{e}$ to be dimensionless,  and (b) becomes the requirement 
\begin{equation}
\langle\psi_{m}, (\hat {g}_{e})^{2}\psi _{m}\rangle - \langle\psi _{m}, \hat {g}_{e} \psi _{m}\rangle^{2} \ll 1. \label{SemiclC}
\end{equation} 
From the beginning, we anticipate that the 4-metric from the initial data set introduces a curvature scale $L(m)$ (actually electric and magnetic curvature scales) from the 3-metric determined by $m$.
\\
\section{Estimate for $\epsilon $ and $a$}

It is important to understand that the finite graphs under study in the semiclassical regime are \emph{embedded graphs}, that is they are (properly) embedded into $\Sigma$ in the presence of the \emph{classical} geometry $m$, which the $\psi _{m}$ are designed to describe.  Hence one can meaningfully speak of the typical proper (coordinate independent) edge length $\epsilon$ of the graph in $\Sigma$ which it inherits from $m$'s classical 3-metric $q_{ab}$. This is in clear distinction to the fully quantum mechanical regime, where there is no such classical geometry into which to embed graphs, and so there is no meaningful $\epsilon$. Hence the typical proper edge length $\epsilon$ is a purely semiclassical quantity.  

Here we summarize the embedding procedure, which is described in fuller detail in Appendix A of \cite{ITP}.
The idea is to take graphs and 2-surfaces that live in $\mathbb{R}^{3}$, called model graphs $\check{\gamma}$ and model surfaces $\check {S}$, and to map them into $\Sigma$ by a (differentiable proper) embedding $X:\mathbb{R}^{3}\to \Sigma$ so that $S=X(\check{S})$ and $\gamma = X(\check{\gamma})$. $X$ may then be viewed as coordinatising $\Sigma$. For simplicity, take $\check{\gamma}$ to be a finite cubic lattice with vertex index $\check{v}\in \mathbb{Z}^{3}$ and inter-vertex edges $\check{e}_{I}(\check{v})$ labelled by a direction index $I=1,2,3$ so that $\check{e}_{I}(\check{v})$ for fixed $I, \check{v}$ is a unit interval parallel to the $I$-th coordinate axis of $\mathbb{R}^{3}$.  Coordinatise $\check{S}$ by $(u,v)\in(-N,N)^2$, where $N$ is a positive integer, so $u,v$ are dimensionless.  $\check{S}$ is the 2-surface $\{ \vec{t}(u,v) | (u,v)\in (-N,N)^2 \}$ ($\vec{t}$ is also dimensionless) and $Y^{a}(u,v) \doteq X^{a}(\vec{t}(u,v)): \check{S}\to S$. Partition $\check{S}$ into maximal, open, connected pieces $\check{S}_k$, whose images $S_k$ under $Y$ are called (fundamental) plaquettes, each lying within a unit coordinate cube of $\mathbb{R}^{3}$, let $\check{v}_k$ be the vertex of that cube, let $(u_k, v_k)\in \check {S}_{k}$, and define $\mu _{k}\doteq \int _{\check{S}_{k}} du dv$. Define $P^{I}_{j}(v) \doteq p^{e_{I}(v)}_{j}(m)$ and also
\begin{equation}
n_{I}(k)\doteq n_i(u_{k}, v_{k}) \doteq \epsilon_{IJK}\, t^{J}_{,u}(u,v)\, t^{K}_{,v}(u,v),
\end{equation}
the normal to $\check{S}_{k}$ in the $I$-th direction, and
\begin{equation}
n^{I}_{a}(\vec{t}) \doteq (1/2) \epsilon _{abc}\, \epsilon^{IJK} X^{b}_{,J}(\vec{t}) \, X^{c}_{,K}(\vec{t}).
\end{equation}

STW first constructed the graph-based kinematic gauge-coherent state $\psi _{m}$ for a single edge, and then
extended this to a coherent state on the entire graph by taking the edgewise tensor product
of states over the graph $\gamma$. This way edges contribute to the graph's overall state independently.  We now turn to the relevant operators.

The so-far arbitrary length $a$ entering the fluxes $p^{e}_{j}$ as well as the edge length $\epsilon $ (arising from the classical geometry $m$ the coherent state describes) that lives on $\gamma\in\Sigma$ are both derived from an optimization procedure which occurs following construction of the $\psi _{m}$.\cite{ITP} Here  we outline that optimization, for further details the reader is referred to Appendix B of   \cite{ITP}.  Like those authors, for the sake of simplifying the mathematics to obtain some simple order of magnitude estimates, we limit the discussion to the Abelian limit $U(1)^{3}$ of $SU(2)$. Because of this restriction, one no longer requires the the system of paths $\rho _{e}(x)$ introduced in the definition of   $p^{e}_{j}$.  The classical kinematic observables are the ``magnetic'' and ``electric'' fluxes for some 2-surface $S\subset\Sigma$ intersecting $\gamma$ (as constructed above), and we assume no pathologies like the ``staircase problem''\cite{GCS} occur generally:
\begin{align}
B_{j}(S) & =\int_{S} F_{j}(x)\\
E_{j}(S) & =\int_{S} \left( *E_{j}\right) (x),
\end{align}
where $F_{j}$ is the curvature 2-form (field strength) of $A^{j}$. The optimization determines $a$ and $\epsilon$ by estimating the classical and quantum contributions to the variances of these two observables.  We can express these observables in terms of the above described embedding, such as:
\begin{equation}
B_{j}(S)=\int_{\check {S}} \mathrm{d}u\,\mathrm{ d}v\, \Big[  n_{a}^{I}(\vec{t}) B_{j}^{a}\left(X(\vec{t})\right)\Big]_{\vec{t}=\vec{t}(u,v)}  n_{I}(u,v),
\end{equation}
where $B^{a}_{j}=(1/2) \epsilon ^{abc}F^{j}_{bc} $ is the gravitational magnetic field.
We then have the classical graph-based approximation functions (Riemann sums) for the above classical operators:

\begin{align}
B_{j,\gamma }(S) &=  \sum _{k} \mu _{k}\, n_{I}(u,v) \Big({\frac{1}{2i}}\Big) \left[ h^{j}_{\alpha ^{I}(v_{k})} - \left( h^{j}_{\alpha ^{I}(v_{k})}\right)^{-1} \right]  \label{BCLSUM}\\
E_{j,\gamma }(S) & =  a^{2}\, \sum _{k} \mu _{k}\, n_{I}(u,v) \, P^{I}_j(v_{k}), \label{ECLSUM}
\end{align}
where $\alpha^{I}(v) = X(\check{\alpha }_{I}(\check{v}))\subset\Sigma$ is the image under $X$ of the model plaquette loop 
\begin{equation}
\check{\alpha}_{I}(\check{v}) = \check{e}^{-1}_{J} \circ \check{e}_{K}(\check{v}+b_{J})^{-1} \circ \check{e}_{J}(\check{v}+b_{K}) \circ \check{e}_{K}(\check{v})
\end{equation}
on $\check\gamma$ in the $JK$ plane with $\epsilon _{IJK}=1$, $v=X(\check{v})$, and $ b_I$ is the standard basis of $\mathbb{R}^{3}$. $h^{j}_{\alpha}$ is the $U(1)$-valued holonomy of $A^{j}$ around the loop $\alpha$.

For $G=U(1)^3$ the complexified phase space point is $g^{j}_{I}(v) = e^{P_{j}^{I}(v)} h^{j}_{e_{I}(v)} \in U(1)$. Then one can replace equations (\ref{BCLSUM}) and 
(\ref{ECLSUM}) by using linear combinations of products of $g^{j}_{I}(v)$, $[g^{j}_{I}(v)]^{-1}$, $[g^{j}_{I}(v)]^{*}$, and $\{[g^{j}_{I}(v)]^{-1}\}^{*}$, where $^{*}$ denotes complex conjugation. These are the classical analogues of  the quantum operators $\hat{g}_{I}(v)$, $[\hat{g}_{I}(v)]^{-1}$, $[\hat{g}_{I}(v)]^{\dag}$, and $\{[\hat{g}_{I}(v)]^{-1}\}^{\dag}$, by which normal-ordered products can be defined. For instance, $g^{j}(v) + \{[g^{j}_{I}(v)]^{-1}\}^{*} = h^{j}_{I}(v) (1+O(\epsilon /a)^{4})$. By using these functions, one can suppress sub-leading terms in quantum expressions to arbitrary order in $(\epsilon /a)$. So the quantum graph-based versions of equations (\ref{BCLSUM}) and (\ref{ECLSUM}) for the kinematic operators are:\\

\begin{align}
\hat{B}_{j,\gamma}(S) = & \sum _{k} \mu _{k} n_{I}(u_{k},v_{k}) : \Big( {\frac{1}{2i}} \Big)
\left[   \hat{h}^{j}_{\alpha_{I}(v_{k})} - \left( \hat{h}^{j}_{\alpha_{I}(v_{k})} \right) ^{-1} \right]  : \label{BQMSUM}\\
\hat{E}_{j,\gamma}(S) = & a^{2} \sum _{k} \mu_{k} n_{I}(u_{k},v_{k}) : \hat{P}^{I}_{j}(v_{k}) : \label{EQMSUM}
\end{align}
where :   : denotes normal ordering after using the above approximation to order $( \epsilon /a)^{2n}$ for any desired positive integral $n$. 

Define the classical error to $\mathscr{O}(m)$  by
 \begin{equation}
\left(\Delta\mathscr{O}\right)_{\mathrm{cl}}(m)\doteq \left|\frac{\mathscr{O}_{\gamma}(m)}{\mathscr{O}(m)} \ -\ 1 \right|
\end{equation}
where $\mathscr{O}$ is the quantity $E_{j}(S)$ or $B_{j}(S)$ above. Then, 

\begin{align}
E_{j}(S)(\Delta E_{j}(S))_{\mathrm{cl}}  & \leq a^{2}\sum _{k} \mu _{k} \Big|  n_{I}(u_{k},v_{k})  \left[ : \hat{P}^{I}_{j}(v_{k}) :\, - \,  P^{I}_{j}(v_{k}) \right] \Big| + \label{ECLASS}\\
& \Big| E_{j}(S) - \sum _{k} \mu _{k} n_{I}(u_{k}, v_{k}) E_{j}(S^{I}(v_{k})) \Big|, \nonumber 
\end{align}
where $S^{I}(v)$ is the image under $X$ of $\check{S}^{I}$ dual to $\check{e}_{I}(\check{v})$. There is an similar expression for $(\Delta B_{j}(S))_{\mathrm{cl}}$.
Here the pair of colons (: :) denote replacement of the functions inside them by the functions of $g^{j}_{I}(v)$ discussed earlier. We analyze the two terms on the RHS separately.
Let $A_S(E)$ be the classical area of $S$ as determined by $E$. Then the first term is bounded above by $A_S(E) (\tilde{E}^{j}_{S})^{n} (\epsilon /a)^{2(n-1)}$ where $\tilde{E}^{j}_{S} $
is the maximum of $| E_j(S')/A_{E}(S') |$ as $S' \subset S$ is varied. 

The second term of the classical error on the RHS of (\ref{ECLASS})  arises because one is making a discrete graph-based approximation $\mathscr{O}_{\gamma}(m)$ (with typical edge length $\epsilon$) to a continuous classical quantity $\mathscr{O}(m)$.   
This difference is the same as that between a Riemann integral and its approximating Riemann sum, given by standard Euler-MacLaurin estimates. Specifically, in one dimension:\begin{align}
\int_{y}^{y+N\epsilon} \,\mathrm{d}x\,F(x) - & \epsilon \left[ \frac{F(y)+F(y+N\epsilon )}{2} + \sum _{k=1}^{N-1} F(y+k\epsilon ) \right] = \\
& \frac{\epsilon ^{3}}{2} \int _{0}^{1}\,\mathrm{ d}t\, \phi (t) \sum _{k=0}^{N-1} F''(y+(k+t)\epsilon ),\nonumber
\end{align}
where $\phi (t) \doteq t^{2}-t$ is the Bernoulli polynomial of second degree, and $F \in C^{2}([y,y+N\epsilon ])$.  Recall $N$ is a natural dimensionless number, and for integrals over $S$, $x, y,$ and $\epsilon$ all have dimensions of length.  For the 2-dimensional case we iterate this expression to obtain:
\begin{align} 
\int _{y_{1}} ^{y_{1} + N_{1}\epsilon} & \, \mathrm {d}x_{1}\,  \int _{y_{2}} ^{y_{2}+N_{2}\epsilon}\,\mathrm{d}x_{2}\, F\big(x_{1}, x_{2}\big) -\big(\mathrm{Riemann\;Sum}\big) = \nonumber \\
& \frac{\epsilon ^{4}}{2} \int _{0}^{1} \,\mathrm{d}t \,\phi (t) \Big[ \sum_{k_{1}=0}^{N_{1}} \sum _{k_{2}=0}^{N_{2}-1} a(k_{1}, k_{2}) F_{22} \big(y_{1}+k_{1}\epsilon , y_{2}
+(k_{2}+t)\epsilon\big) + \nonumber\\
& \sum _{k_{1}=0}^{N_{1}-1} \sum _{k_{2}=0}^{N_2} b(k_{1}, k_{2}) F_{11}\big(y_{1}+(k_{1}+t)\epsilon , y_{2}+k_{2}\epsilon\big) \Big] + \\
& \frac {\epsilon^{6}}{4} \int_{0}^{1}\,\mathrm{d}t_{1}\, \int_{0}^{1}\,\mathrm{d}t_{2}\, \phi(t_{1})\phi(t_{2})\, \sum_{k_{1}=0}^{N_{1}-1} \sum_{k_{2}=0}^{N_{2}-1} F_{1122}\big(
y_{1}+(k_{1}+t_{1})\epsilon ,y_{2}+(k_{2}+t_{2})\epsilon\big), \nonumber
\end{align}
where $a(k_{1}, k_{2}) = 1/2$ for $k_{1}=0, N_{1}$, and $ a=1$ otherwise, $b(k_{1},k_{2}) = 1/2$ for $k_{2}=0, N_{2}$ and $b=1$ otherwise, $F_{ij}= \partial _{i}\partial_{j} F$,
and $F_{ijkl}=\partial_{i}\partial_{j}\partial_{k}\partial_{l} F$. Define
\begin{align}
& \qquad\qquad\qquad\qquad\left(\frac{1}{L_{F}(j,S)} \right)^{2} \doteq  \Bigg| \left(\int_{S} \,\mathrm{d}^{2}x\, F \right) ^{-1} \frac{\epsilon ^{2}}{2} \int _{0}^{1} \,\mathrm{d}t\, \phi (t) \, \times \\
  &  \Big[  \sum _{k_{1}=0}^{N_{1}} \sum _{k_{2}=0}^{N_{2}-1} a(k_{1},
k_{2}) F_{22} (y_{1}+k_{1}\epsilon , y_{2}+(k_{2}+t)\epsilon) + 
 \sum_{k_{1}=0}^{N_{1}-1} \sum _{k_{2}=0}^{N_{2}} b(k_{1},k_{2}) F_{11} (y_{1} +(k_{1}+t)\epsilon, y_{2}+k_{2}\epsilon)
\Big] \Bigg| \nonumber
\end{align}
and
\begin{align}
\left(\frac{1}{\xi_{F}(j,S)} \right) ^{4} \doteq & \Big| \left( \int _{S} \,\mathrm{d}^{2}x\, F\right)^{-1} \frac{\epsilon^{2}}{4} \int _{0}^{1} \,\mathrm{d}t_{1} \,\int _{0}^{1} \,\mathrm{d}t_{2} \,
\phi (t_{1}) \phi(t_{2}) \, \times \\
& \sum _{k_{1}=0} ^{N_{1}-1} \sum
_{k_{2}=0}^{N_{2}-1} F_{1122}(y_{1} + (k_{1}+t_{1})\epsilon, y_{2} + (k_{2}+t_{2})\epsilon)\Big| \nonumber ,
\end{align}
where the quantities on both LHS's acquire a 2-surface $S$-dependence as well as a Lie generator index $j$-dependence from $E_{j}(S)$ 
and $B_{j}(S)$, respectively. Since $|\phi (t) | \le 1/4$ on $t\in [0,1]$, and denoting the 2-dimensional Riemann sum by $\mathrm{R\_Sum}$, we find the bound
\begin{equation}
\Delta _{2} (F) \doteq \left| \int _{S} d^{2}x F(x) - \mathrm{R\_Sum} \right| \le 
\left[\left( \frac{\epsilon}{L_{F}(j,S)} \right)^{2}+
\left( \frac{\epsilon}{\xi_{F}(j,S)} \right) ^{4}\right]  \left| \int _{S} \,\mathrm{d}^{2}x\, F \right| .
\end{equation}
From the definition of the length $L_{F}(j,S)$ one has
\begin{equation}
L_{F}(j,S)^{-2} \lesssim \frac{1}{8} \left| \int _{S} \,\mathrm{d}^{2}x\, F \right| ^{-1} \left| \int _{S} \,\mathrm{d}^{2}x\, \left( F_{11}+F_{22}\right)\right|,
\end{equation}
so from this inequality (or directly from its definition) one sees that $L_{F}(j,S)^{-2}$ is a $(j,S)$-dependent weighted average of inverse-squared bending lengths of $F(x)$ over the 2-surface $S$. To eliminate consequences of the $(j,S)$-dependence, evaluate the minimum
\begin{equation}
L_{F}\doteq L_{F}(m) \doteq \min _{j,S} L_{F}(j,S),
\end{equation}
where the minimum is taken as $j$ and $S$ are varied over their respective ranges while keeping the classical phase space point $m$ fixed. I.e., one ``throws'' all compact connected smooth 2-surfaces $S$ of suitably simple topology
 ``into'' the fixed 3-geometry of $\Sigma$ that is just the given classical 3-geometry $m$.
The length $L(F)$ can be, respectively, no smaller than the \emph{minimum} of the electric (respectively, magnetic) 3-curvature radius that occurs within the classical 3-geometry
$m$, and $L_{F}(j,S)^{-2}$ is then bounded above by the inverse-square of the appropriate minimum 3-curvature radius. This can be seen just by choosing the 2-surface S to closely surround the locations in the 3-geometry $m$ where those minimum curvature radii are attained. Alternatively, since $L_{F}(j,S)^{-2}$ is an \emph{average} over $S$ of inverse-squared bending lengths of $*E_{j}(x)$ or the $j$-th Lie component of the field strength $F_{j}(x)$, and then defining $L_{E}$ and $L_{B}$ to be the minimum electric and magnetic curvature lengths in $m$, one has immediately has that
$L_{E}(j,S)^{-2} \le L_{E}^{-2}$ and $L_{B}(j,S)^{-2} \le L_{B}^{-2}$.
Similarly, one finds
\begin{equation}
\xi_{F}(j,S)^{-4} \lesssim \left(\frac{1}{4}\right)^{3} \left| \int_{S} \,\mathrm{d}^{2} x \, F \right| ^{-1} \left| \int_{S} \,\mathrm{d}^{2}x\, F_{1122} \right| ,
\end{equation}
so (or directly from its definition) $\xi_{F}(j,S)^{-4}$ is another average of inverse quartic bending lengths, and one correspondingly sets 
\begin{equation}
\xi_{F}\doteq \xi_{F}(m)\doteq \min _{j,S} \xi_{F}(j,S) \thicksim O(L_{F}).
\end{equation}
Therefore, to leading order in $\epsilon /L_{F}$, one has the bound on the error of the Riemann sum approximation to the 2-dimensional Riemann integral:
\begin{equation}
\left| \Delta_{2}(F) \right| \le\left[\left(\epsilon/L_{F}\right)^{2} + O(\epsilon/L_{F})^4\right] \left| \int_{S} d^{2}x \, F \right| .
\end{equation}
Assume (as will be subsequently verified, see equations (\ref{AEST}) and (\ref{EPSIEST}) below) that $\epsilon \ll a, L_{E}, L_{B}$, and take $n$ to be sufficiently large so that the first term in the classical errors for $E$ and $B$ are dominated by the second term coming from the Riemann sum error (see equation (\ref{ECLASS}) above and its following discussion), one then obtains the classical errors:
\begin{equation}
\left(\Delta E_{j}(S)\right)_{\mathrm{cl}} \le \left(\epsilon/L_{E}\right)^{2} \label{ERRECL}
\end{equation}
and
\begin{equation}
\left(\Delta B_{j}(S)\right)_{\mathrm{cl}} \le \left(\epsilon/L_{B}\right)^{2}.\label{ERRBCL}
\end{equation}

These estimates differ on their RHS's from that of STW, who have 
$(\epsilon / L_{E})^{4}$ and $ (\epsilon / L_{B})^{4}$, respectively.\cite{ITP} Carefully iterating the Euler-MacLaurin formula from 1- to 2-dimensional integrals, turns out \emph{not} to increase the relative accuracy of the Riemann sum approximation to the corresponding Riemann integral. \\

Next, we turn to the quantum contributions to the variances of $E_{j}(S)$ and $B_{j}(S)$. Both $\hat{B}_{j,\gamma}(S)$ and $\hat{E}_{j,\gamma}(S)$ are of the form 
$\hat{ \mathscr{O}} = \sum _{k} \hat {\mathscr {O}}_{k}$, where the $k$-sum is over fundamental plaquettes that comprise the surface $S$. 
For $\hat{E}$ the $\hat{\mathscr{O}}_{k}$ mutually commute, but this does
not occur for $\hat{B}$ where neighboring terms (plaquettes) are non-commuting. However, the number of such non-commuting terms is of the same order as the number of $\hat{\mathscr{O}}_{k}$, so if we just wish to estimate orders of magnitude of the magnetic quantum fluctuations,  their contribution is a factor of order unity, which we may ignore. So taking all the $\hat{\mathscr{O}}_{k}$ to mutually commute, for fluctuations in the coherent state $\psi_{\hat{\gamma},\epsilon ,X,m}$ we have:
\begin{align}
\langle\hat{\mathscr{O}}\rangle ^{2}  \left(\Delta\mathscr{O}\right)(m)_{\mathrm{qnt}}^{2} & = 
\sum _{k} \left[ \langle\hat{\mathscr{O}}_{k}^{2}\rangle - \langle\hat{\mathscr{O}}_{k}\rangle^{2} \right]   \nonumber    \\
& = N(\epsilon ,S) \left[ \left| \frac{\partial\mathscr{O}_{k}}{\partial h_{e(k)}} \right| ^{2} \mathrm{Var} \left(\hat{h}_{e(k)}\right) + \left| \frac{\partial\mathscr{O}_{k}}{\partial p_{e(k)}} \right| ^{2} \mathrm{Var} \Big(\hat{p}_{e(k)}\Big) \right]  \label{QMERR},
\end{align}
where $ N(\epsilon , S)$ is the number of quantum mechanically \emph{independently} fluctuating plaquettes $k$, and Var denotes the quantum mechanical variance of its operator-valued argument in the coherent state $\psi_{\hat{\gamma},\epsilon ,X,m}$.  One might choose $N(\epsilon, S)=A_{E}(S)/\epsilon^{2}$, where $A_{E}(S)$ is the classically determined area of $S$ from the flux variables $E_{j}$, and use the Law of Large Numbers for the fluctuators (plaquettes or edges). But for later insight, we retain $N(\epsilon ,S)$ as a separate variable. Using the results from \cite{GCS} one has:
\begin{equation}
\mathrm{Var} (\hat{h}_e) = \mathrm{Var} (\hat{p}_e) = t,
\end{equation}
where $t\doteq(L_{P}/a)^{2}\ll 1$ is the classicality parameter (the inequality is verified later, see equations (\ref{AEST}) and (\ref{EPSIEST}) below). 
Next we utilize the classical expressions (\ref{BCLSUM}) and (\ref{ECLSUM})
for $B_{j, \gamma}(S)$ and $E_{j, \gamma}(S)$, respectively, to evaluate the partial derivatives on the RHS of  (\ref{QMERR}). For $\mathscr{O}=E$ with Abelian $G=U(1)^3$ the $h_e$ factors drop out of the definition of $p_{e}^{j}$ in (\ref{PDEFN}), and the non-vanishing term is
\begin{equation}
\left| \frac{\partial\mathscr{O}_{k}}{\partial p_{e(k)}} \right| = a^{2},
\end{equation}
so
\begin{equation}
\langle \hat{E}(S)\rangle ^{2} (\Delta E)^{2}_{\mathrm{qnt}}(m) = N(\epsilon, S)\, t \,a^{4}  \label{QMEERR}.
\end{equation}
As a check: both sides have dimensions $E^2$ or area$^2$.
When
 \begin{equation}
 \mathscr{O}=B_{j,\gamma}(S)\doteq \sum_{k} B_{k}^{j}  = \sum _{k} \mu _{k} n_{I}(u_{k}, v_{k}) \Big(1/2i\Big) \left[ h^{j}_{\alpha ^{I}(v_{k})} - (h^{j}_{\alpha ^{I}(v_{k})})^{-1} \right]
 \end{equation}
 and $G=U(1)^{3}$, one has
 \begin{equation}
  h^{j}_{\alpha ^{I}(v_{k})} =\exp{ \left[ i \int _{\alpha^{I}(v_{k})} A^{j}\right]} \doteq \exp {\left[ i\Phi ^{j}_{\alpha^{I}(v_{k})} \right] } \in U(1).
  \end{equation}
  (Recall the connection $A^{j}$ is real-valued, and $h^{j}$ is U(1)-valued.) Consequently,
  \begin{align}
  \left| \frac{\partial B_{k} ^{j'}} {\partial h^{j}_{\alpha (v_{k})}}  \right| & = \left| \delta^{j'}_{j} \mu _{k} n_{I}(u_{k}, v_{k}) \frac{\partial}{\partial h^{j}_{\alpha (v_{k})}} \sin\left(\Phi^{j}_
  {\alpha^{I}(v_{k})}\right) \right| \nonumber \\ 
  & = \left| \delta^{j'}_{j} \mu _{k} n_{I}(u_{k}, v_{k}) \cos\left(\Phi^{j}_{\alpha^{I}(v_{k})}\right) \left[ \frac{\partial h^{j}_{\alpha (v_{k})}}  {\partial\Phi^{j}_{\alpha (v_{k})}} \right]^{-1} \right| \nonumber \\
  & = \left| \delta^{j'}_{j} \mu _{k} n_{I}(u_{k}, v_{k}) \cos\left(\Phi^{j}_{\alpha^{I}(v_{k})}\right) \exp \left[ -i \Phi^{j}_{\alpha^{I}(v_{k})}\right] (-i) \right|.
  \end{align}
However, $ | \mu _{k} n_{I}(u_{k}, v_{k}) | \le 1$, because they live on the model manifold $\check{S}$. Additionally, henceforth specializing to weak gravitational fields such as those prevailing in our solar system,
 $|\Phi_{k}|\simeq (\epsilon / L_{B})^{2} \ll 1$ and $\cos \Phi_{k} \simeq 1$. So under these conditions, $\left| \partial B^{j'}_{k}/\partial h^{j}_{\alpha(k)} \right| \simeq \delta_{j}^{j'}$. 
 Summarizing,
  \begin{equation}
  \langle\hat{B}(S)\rangle^{2} (\Delta\hat{B})^{2}_{\mathrm{qnt}}(m) \lesssim N(\epsilon, S) \,  t,  \label{QMBERR}
  \end{equation}
and both sides are dimensionless. \\

 By adding the classical and quantum variances (assumed to be independent) for both of $E(S)$ and $B(S)$  we may then estimate their corresponding total variances.
 For the total variance of $E(S)$, we use equations (\ref{ERRECL}) and (\ref{QMEERR}) to find
 \begin{align}
 \left[\Delta E_{j}(S)\right]^{2}_{\mathrm{Total}} (m) & = \left[\Delta E_{j}(S)\right]^{2}_{\mathrm{cl}} (m) + \left[\Delta E_{j}(S)\right]^{2}_{\mathrm{qnt}} (m) \nonumber \\
 & \lesssim\left(\frac{\epsilon}{L_{E}}\right)^{4} + \frac{N(\epsilon, S)\, t \, a^{4}}{E_{j}(S)^{2}}.
 \end{align}
 Similarly, for the total variance of $B(S)$ equations (\ref{ERRBCL}) and (\ref{QMBERR}) yield
 \begin{align}
 \left[\Delta B_{j}(S)\right]^{2}_{\mathrm{Total}} (m) & = \left[\Delta B_{j}(S)\right]^{2}_{\mathrm{cl}} (m) + \left[\Delta B_{j}(S)\right]^{2}_{\mathrm{qnt}} (m) \nonumber \\
 & \lesssim \left(\frac{\epsilon}{L_{B}}\right)^{4} + \frac{N(\epsilon, S)\, t}{B_{j}(S)^{2}}.
 \end{align}
For generic weak classical gravitational fields, like those considered here, one expects $L_{E}\simeq L_{B}$, since both (minimum) 3-curvatures arise from (the minimum of) the same 4-curvature (Riemann) tensor describing $m$.
So we may set $L_{E}\simeq L_{B} \simeq L_{C}$, the \emph{minimum} curvature radius of the classical geometry $m$. 
This expectation is exact for spacetime geometries with vanishing extrinsic curvature $K$, then the coordinate dependent slicing of spacetime
does not affect $L_E$ and $L_B$. For the estimates to be made below we use the Schwarzschild geometry, for which $K=0$.
Further approximating
$B_{j}(S)\simeq A_{E}(S)/L_{C}^{2}$ and $\sqrt{E_{j}(S)^{2}} \simeq A_{E}(S)$, one obtains
\begin{align}
\left[\Delta E_{j}(S)\right]^{2}_{\mathrm{Total}}(m) \lesssim \frac{\epsilon ^{4}}{L_{C}^{4}} \, +\, (tN) \frac {a^{4}}{E_{j}(S)^{2}} & \simeq\frac{\epsilon^{4}}{L_{C}^{4}} \, +\, \frac{L_{P}^{2}a^{2}}{A_{E}(S)^{2}} N(\epsilon , S)\label{Efluc}\\
\left[\Delta B_{j}(S)\right]^{2}_{\mathrm{Total}}(m) \lesssim \frac{\epsilon ^{4}}{L_{C}^{4}} \, +\, (tN) \frac {1}{B_{j}(S)^{2}} & \simeq\frac{\epsilon^{4}}{L_{C}^{4}} \, +\, \frac{L_{P}^{2}L_{C}^{4}}{a^{2}A_{E}(S)^{2}} N(\epsilon , S).\label{Bfluc}
\end{align}
Imposing balanced variances: $\left(\Delta E_{j}(S)\right)^{2} = \left(\Delta B_{j}(S)\right)^{2}$ implies $a=L_C$, \emph{independent} of both the choice of $N(\epsilon ,S)$ as well as the value of
 $\epsilon$.  This simple result for $a$ is intuitively gratifying. After all, the classicality parameter $t=(L_{P}/a)^{2}$ is supposed to measure robustly just how ``classical'' the geometry 
 $m$ is, and the minimum curvature length $L_C$ is the first scale coming to mind that so characterises $m$. $L_C$ relative to the quantum scale $L_P$ would then 
 be a reasonable first guess for a quantity like 
 ``classicality.''  This result then provides some confidence that the total variance optimization calculation is on the right track.\\

To obtain an estimate for $\epsilon$ one then additionally minimizes either the total $\Delta E_{j}(S)$ or the total $\Delta B_{j}(S)$ with respect to $\epsilon$ while setting $a=L_C$. 
Notice that for this to yield a
non-vanishing optimal value for $\epsilon$ requires $N(\epsilon ,S)$ to be $\epsilon $ dependent; such as $N=A_{E}(S)/\epsilon^{2}$, which assumes the elemental 
surfaces $S_{e}$ to fluctuate independently.  If we take that Ansatz, then minimizing $\left(\Delta E _{j}\right)^{2}$ requires us to minimize the expression\\
\begin{equation}
\frac{\epsilon ^{4}}{L_{C}^{4}} \ +\ \frac{L_{P}^{2}L_C^{2}}{\epsilon ^{2} L_{S} ^{2}}
\end{equation}
with respect to $\epsilon $, where we have written $A_E(S)\doteq L_{S} ^{2}$. Self-consistency in the semiclassical regime requires $L_{P}\ll\epsilon\lesssim L_{S}\lesssim L_{C}$. Hence there are two extremes for the range of values for $\epsilon$, corresponding to the allowed range of $L_{S}$ over which  measurements of geometric operators like $E_{j}(S)$ and $B_{j}(S)$ are performed. We turn to each of those now.\\

Case (I): Here $\epsilon\simeq L_{S}$, and the geometric measurements operate on the scale of $\epsilon$, namely over a typical single edge proper length. This is the same as setting $N(\epsilon, S)\simeq 1$. Denoting  $\epsilon_{1} \doteq\epsilon$ for this extreme value, we need to minimize \\
\begin{equation}
\frac{\epsilon _{1}^{4}}{L_{C}^{4}}  + \frac{L_{P}^{2}L_C^{2}}{\epsilon _{1}^{4}}
\end{equation}
for $\epsilon_1$, which occurs when the two terms are approximately equal. This yields the estimate
\begin{equation}
\epsilon_{1}\simeq L_{P}^{1/4}L_{C}^{3/4}.
\end{equation}
As an aside we note STW \cite{ITP} obtained $\epsilon_{STW}\simeq L_{P}^{1/6}L_{C}^{5/6}$, differing from this expression due to the way they iterated the
Euler-Maclaurin estimation from a single to a double integral. \\

Case (II): Here N takes its maximum value or $L_{S}\simeq L_{C}$. In this case, geometric operators measure scales all the way up to the classical curvature length $L_C$.
Writing this solution as $\epsilon_{M}$, it must minimize\\
\begin{equation}
\frac{\epsilon _{M}^{4}}{L_{C}^{4}} \ +\ \frac{L_P^{2}}{\epsilon _{M}^{2}},\\
\end{equation}
from which 
\begin{equation}
\epsilon_{M}\simeq L_{P}^{1/3} L_{C}^{2/3} = \left(\frac{L_{P}}{L_{C}}\right)^{(1/12)} \epsilon_{1}<\epsilon_{1}.
\end{equation}
\\
Summarizing: 
\begin{equation}
(L_{P}/L_{C})^{1/12} \epsilon_{1}\lesssim\epsilon\lesssim\epsilon_{1}\simeq L_{P}^{1/4} L_{C}^{3/4}\label{EPSIEST}
\end{equation}
and 
\begin{equation}
a\simeq L_{C}. \label{AEST} 
\end{equation}
From these we see that the total variance optimization calculation is indeed self-consistent: $L_{P} \ll \epsilon \ll a\simeq L_{C}$.

But just how long is this ``new mesoscopic'' scale $\epsilon$? More specifically, how large \emph{in meters} are these semiclassical states of space expected to be right outside our windows, here on planet Earth?  We may compute the curvature length $R_{m}(x)$ at the Earth's surface by taking our Blue Planet to fairly good approximation to be a sphere of radius $R_E$ and mass $M_E$, neglecting for the moment the effects of other bodies in the Solar system like the Moon, Sun, and other planets. We can then estimate $L_{C}=\min _{x} R_{m}(x)$ outside our windows
by using expressions from well-known textbooks on classical gravitation \cite{MTW_LC} as:
\begin{equation}
L_{C}(\mathrm{window}) \simeq\left(\frac{c^{2}R_{E}^{3}}{GM_{E}}\right)^{1/2}\simeq 2.41\cdot 10^{11} \; \mathrm{m}.\label{LCEQN}
\end{equation}
Notice $R_{m}(x)$ precisely attains its minimum just at the Earth's surface $|x|=R_E$, and then is constant as one drills into the Earth, assuming the planet to have a constant mass density.
Using $L_{P}=(\hbar G/c^{3})^{1/2}\simeq 1.6\cdot 10^{-35}\ \mathrm{m}$, we estimate 
\begin{equation}
\epsilon_{M}\simeq 98\ \mu\mathrm{m} \lesssim \epsilon\lesssim\epsilon_{1}\simeq 0.69\  \mathrm{m}.\\
\end{equation}
This is truly an astonishingly long semiclassical gravitational length scale! It implies the existence of (kinematic) quantum coherent states of the gravitational field on Earth having sizes in the macroscopic 100 $\mu$m to 1 meter range,  potentially rivaling superconductors' quantum coherence over ``miles of dirty lead wire.'' One can also estimate $\epsilon_{STW}$ on Earth,
and it comes out to just below 5 km, yet longer. Additionally, using the equations of geodesic deviation\cite{MTW_GD} one can calculate the size of the solar and lunar tidal modulations of $L_C$ on Earth, and they affect $\epsilon$ only roughly at the 10 ppm level. We now turn to the possible experimental consequences of a typical graph edge length having this magnitude.\\
\section{Physical Consequences of the Length Scale $\epsilon $ }

In this section we present some further arguments and state the results of some straightforward computations, whose details are omitted since they are  applications
of previous theory to this specific regime. \\

There will be quantum fluctuations of the  flux and holonomy variables about the optimized edge length $\epsilon$. These may be pictured
as fluctuations of the geometry such that the length $\epsilon$ fluctuates to $\epsilon + \delta\epsilon$. Since such a fluctuation is quantum in origin, it is off-shell in nature, i.e., the fluctuations themselves are not solutions of the Einstein field equations. The probability of the the fluctuation $\delta\epsilon$ can be calculated from the normalized single-edge overlap $i_{t} (g,g')$ of kinematic gauge coherent states originally derived by Thiemann.\cite{GCS} In the full non-Abelian case, $g$ is the element of $SL(2,\mathbb{C}) = SU(2)^{\mathbb{C}}$ corresponding to the classical edge-based pair of holonomy
$h_e$ and flux $p_{j}^{e}$ geometric variables giving edge length $\epsilon$, and $g'$ corresponds to a similar complexified gauge group element for length $\epsilon + \delta\epsilon$. Assuming variables in the range found near the Earth's surface, several pages of straightforward but not particularly illuminating
calculation lead to the simple Gaussian \\
\begin{align}
i_{t}(g,g')\doteq i_{t}(\epsilon , \delta\epsilon)  &  \propto\exp \Big[ -  \left(\frac{\delta\epsilon}{L_{I}}\right) ^{2}   -  \left(  \frac{\delta\epsilon}{L_{II}}\right)^{2}\Big]   \\
L_{I}  &  =(1/2) L_{P} \\
L_{II} &  =(1/2) L_{P} \left(\frac{L_{C}}{L_{P}}\right)^{1/4}\gg L_{I},
\end{align}
where a standard Gaussian normalization factor has been omitted for clarity. $L_{I}$ predominately arises from fluctuations of holonomies, and is at the Planck scale. $L_{II}$ originates from fluctuations in fluxes or areas, and on the Earth's surface takes the value
$2.8\cdot 10^{-24}\; \mathrm{m}$. This is still a short length scale, but the first Planck scale term in the exponential will suppress it. Consequently, for Earth bound laboratories the 
edge length fluctuations have root mean square amplitude firmly at the Planck scale $L_{P}$. \\

The above analysis really describes an independently fluctuating grain of \emph{space} of typical size $\epsilon$. Can it be used to say something more about \emph{spacetime}?   If we can use basic causality arguments familiar from classical spacetime in the semiclassical regime, the characteristic size $\epsilon$ would imply a characteristic upper roll-off frequency $\omega_{0}$ in the fluctuation spectrum $S(\omega)$,
given roughly by the inverse of a grain's light transit time $\tau_{\epsilon}$: $\omega_{0} \simeq (1/\tau_{\epsilon}) \simeq c/\epsilon$. For the range of values of $\epsilon$
on Earth given earlier, this roll-off would lie between 0.43 GHz and 3.0 THz.
This information is sufficient to roughly estimate
the normalized spectrum of the spacetime strain $h\doteq \delta\epsilon/\epsilon$ from the fluctuations of a single grain as
\begin{equation}
S_{h}(\omega)\simeq \left(2/\pi\right)^{1/2} h_{1} \left[1+(\omega \tau_{\epsilon})^{2}\right] ^{-1/2} \sqrt {\tau_{\epsilon}}\quad  (\mathrm{Hz}^{-1/2}),  \\
\end{equation}
with $h_{1}\simeq L_{P}/\epsilon$ being the root mean variance of the strain fluctuations for one edge. It is important to reiterate that these fluctuations are off-shell, and so cannot be thought of as propagating gravitational waves. Also, LQG theory has not yet advanced to the point where either causality or 4-metric/spacetime strain fluctuations are explicitly included or calculated. Nevertheless, it is physically plausible that the kind of length fluctuations described by the above expression for $i_t(\epsilon,\delta\epsilon)$  might be detectable by experiments sensitive to spacetime strain fluctuations. In fact, the use of such techniques generally to test theories of quantum gravity was suggested about a decade ago. For some reviews see \cite{Amelino_Camelia}. \\

The experimental methods currently underway use two different approaches to detect small
spacetime strains. The first uses optical interferometry over long baselines (several km for LIGO), and are optimally sensitive to the 100's of Hz spectral window. Advanced LIGO
is expected to have a strain noise floor of $S_{h} (\mathrm{LIGO\; floor})\simeq (3.8-4.0)\cdot 10^{-24}\; \mathrm{Hz}^{-1/2}$ in this frequency band.\cite{LIGO}
 LIGO type interferometers with an effective photon round trip distance $\Lambda$ of 10 km for $\epsilon \simeq \epsilon _{1} \simeq 0.7\;\mathrm{m}$ will   be bathed by a strain noise  density $S_{h}(\epsilon_{1})\simeq (1.1\cdot10^{-37}\;\mathrm {Hz}^{-1/2}) (\Lambda/10\,\mathrm{km})^{1/2}$; where it has been assumed that the strain spectrum scales proportionately to $(L_{P}/\epsilon)\, (\Lambda/\epsilon)^{1/2}$. For $\epsilon\simeq\epsilon_{M}\simeq 100\,\mu\mathrm{m}$, the noise spectral density is 
$S_{h}(\epsilon_{M})\simeq (7.3\cdot 10^{-34}\;\mathrm{Hz}^{-1/2}) (\Lambda/10\,\mathrm{km})^{1/2}$. These fluctuations are  therefore about 10 orders of magnitude too small to be detectible by Advanced LIGO.\\

There are alternative non-interferometric experimental methods, 
such as Li-Baker, that are sensitive to a varying 4-geometry such as propagating gravitational waves by relying on a classical resonant propagating gravitational-electromagnetic wave interaction.\cite{Li_Baker} In order for these to be sensitive to length fluctuations $\delta\epsilon\simeq L_{P}$, one has to make the additional assumption that the quantum fluctuations have an appreciable (near unity) projection onto off-shell 4-metric fluctuations. The off-shell, non-propagating, stochastic nature of the 4-metric fluctuations implies that the resonance enhanced detection in the few GHz microwave range that normally acts for propagating
classical gravitational waves no longer applies.  There will consequently be a reduction in experimental sensitivity that can be estimated by setting the enhancement
factor ($k\Lambda\simeq 630$ for microwave wavenumber $k$) to unity. However, Li-Baker detectors have the virtue of just being able to enter the GHz regime where the fluctuations' high frequency roll-off would be starting to occur.
From literature about the Li-Baker technique\cite{Li_Baker}, if we take the operating frequency to be that of the current prototype, 5 GHz, and an electromagnetic-gravitational interaction length $\Lambda$ of 6 m, one finds for long $\epsilon$: $S_{h}(\epsilon_{1}\simeq 0.7\;\mathrm{m})\simeq 5.6\cdot10^{-40}\;\mathrm{Hz}^{-1/2}$, and for short $\epsilon$ : $S_{h}(\epsilon_{M}\simeq 100\;\mu\mathrm{m})\simeq 1.1\cdot10^{-34}\; \mathrm{Hz}^{-1/2}$. This is about $5.5\cdot10^4$ times smaller than the minimum detectible stochastic (non-resonant) Li-Baker sensitivity of $6.3\cdot 10^{-30}\;\mathrm{Hz}^{-1/2}$.
Again, it has been assumed that the strain spectrum scales proportionately to $(L_{P}/\epsilon)\, (\Lambda/\epsilon)^{1/2}$. 
Hence the near term prospects of observing such strain or metric fluctuations arising from $\delta\epsilon$ are remote for the Li-Baker type detectors as well. \\

In summary, in spite of the macroscopic size of $\epsilon$ itself, the associated quantum fluctuations still lie beyond the reach of current experimental technology. Taken from another viewpoint, remarkably no experimental spacetime strain measurements so far contradict the theoretical possibility of states of space having the range of sizes $100\, \mu\mathrm{m}\, -\, 0.7\, \mathrm{m}$ on the Earth's surface.\\

\section{Deeper Problems with $\epsilon$}

There is, however, a more fundamental difficulty with $\epsilon\simeq 100\, \mu \mathrm{m}-0.7\, \mathrm{m}$. That occurs when one attempts to construct an effective semiclassical Hamiltonian for LQG coupled to other fields (``matter''), such as the Klein-Gordon or electromagnetic (Maxwell) fields, which have a characteristic wavelength 
$\lambda$.\cite{Sahl_Thesis} This might be the Compton wavelength of the Klein-Gordon field, or just the photon wavelength. For example,  for optical photons $\lambda$ is roughly $0.3-0.8\,\mu\mathrm{m}$. To analyze this specific problem, the Hamiltonian is typically written as a spatial integral of a product of fields describing matter and the gravitational degrees of freedom, such as\\
\begin{equation}
\mathscr{H}_{EM} = \left(\frac{1}{2Q_{EM}}\right) \int N(x) \left(\frac{q_{ab}}{[\det(q_{ab})]^{1/2}}\right)\left[\underline{E}^{a}\underline{E}^{b} + \epsilon^{acd} \underline{A}_{d,c} \epsilon^{bef} \underline{A}_{f,e}\right]  \mathrm{d} ^{3}x.
\end{equation}
Here $Q_{EM}$ is a coupling constant, $N(x)$ is an ADM lapse-like smearing function, $\underline{A}_{a}$ are the spatial components of the electromagnetic $U(1)$ gauge potential one-form (pull-back of the $U(1)$-bundle connection one-form), and $\underline{E}^{a}$ are the EM electric field components.
One then manipulates this expression into a form where one can evaluate the expectation value of the gravitational degrees of freedom in the coherent states, usually by expressing the the gravitational factors in terms of the volume operator, that acts in a known way at the graph's vertices. However, for this to suitably regulate quantities like $\mathscr{H}_{EM}$ and obtain an effective Hamiltonian for classical electromagnetism requires a hierarchy of length scales:
\begin{equation}
L_{P}\ll \epsilon \ll \lambda \ll L_{C}.
\end{equation}
This set of inequalities is important when one makes a spatial gradient expansion in the EM electric and magnetic fields around the graph vertices and retains the leading order terms. So if $\epsilon$ on Earth is longer than $100\,\mu\mathrm{m}$, then it is not possible to to carry out such an expansion for optical electromagnetic radiation, where $\lambda$ is of submicron size. Of course, the problem is even worse for
X- or $\gamma$- radiation; or away from massive astrophysical objects in interstellar or intergalactic space where $L_{C}$ (and therefore $\epsilon$) becomes even larger. To ensure that the semiclassical effects of LQG on propagating radiation are small corrections to the classical Maxwell equations requires $L_{P} \ll \epsilon \ll \lambda$, otherwise the regulation fails in the semiclassical regime. There may be other length scales associated with graphical LQG, such as the graph isotropy or homogeneity scales\cite{Sahl_Thesis} that also play a role in the regularization, but here the focus is only on $\epsilon$.\\

What went awry in the relatively straightforward estimate for $\epsilon$? If we return to equations (\ref{Efluc}) and (\ref{Bfluc}) for the simplified gravitational gauge group $U(1)^{3}$, we see that for operators defined by a surface $S$, the quantity $N(\epsilon, S)$ enters as the number of independent fluctuating edges or coherent states contributing to those operators.  By setting $N(\epsilon, S) = A_{E}(S)/\epsilon^{2}$, one assumes the coherent state for the entire graph is a tensor product of independently fluctuating edges (or dual elemental plaquettes). If the state for the whole graph were \emph{not a tensor product}, so $N(\epsilon , S)$ were independent of $\epsilon$, then optimization would lead to $\epsilon \to 0$, minimizing only the classical variance. Of course, $\epsilon$ never  actually approaches zero because in LQG edges cannot become much smaller than $L_{P}$. For example, one could choose the entire surface $S$ to be a single quantum coherent fluctuator by constructing a collective coherent state for it as a \emph{non-tensor product} combination of its component elemental surfaces (or edges that transversely intersect $S$), that would make $N(\epsilon, S) =1$. Recently
Oriti, Pereira, and Sindoni (OPS) have made a proposal \cite{OPS} towards constructing such collective coherent states by using the non-commutative ($\star -$product) flux representation\cite{Flux_Repn} in Lie algebra variables $ \{x_{e}\}$, which may be written as \\
\begin{equation}
\Psi _{\vec {E}(S), \vec{\Phi}(S)}\left(\{x_{e}\}\right) = \bigotimes _{S\in\mathscr{S}} K^{t} \left( \sum _{I(S)} x_{I} \, - \, E(S)\right) \star \exp {\left[\left(\frac{i}{t}\right) \Phi(S) \cdot \left( \sum _{I(S)} x_{I}\right)\right]} \label{OPSSTATE}
\end{equation}
(\cite{OPS} equation (26)). Here $\mathscr{S}=\{S\}$ could be a "complete'' set of collective observables  (embedded surfaces $S$); i.e., such that each embedded graph edge $e$ intersects exactly one $S\in\mathscr{S}$. $ I(S)$ labels all the edges intersecting a given $S$. The over-arrow denotes a vector over the members of $\mathscr{S}$. Here $K^{t}$  is the heat kernel (Gaussian) in the flux representation, and the new collective ``coarse grained'' variables $E(S)$ and $ \Phi(S)$ are defined in OPS. When OPS introduced these coherent states for collective variables, there was no complete collection of surfaces $\mathscr{S}$ and no corresponding complete set of pairs of collective variables. Instead they multiplied the collective part of the state by a state  $\Psi_{\mathrm{Non-coll}} $ describing the residual non-coarse grained (non-collective) degrees of freedom that are needed to specify the classical 3-geometry $m$. Such a factor has been omitted in equation (\ref{OPSSTATE}), and it should be re-instated if the collection $\mathscr{S}$ is insufficient to specify $m$. The choice of $\mathscr{S}$ critically determines the structure of the collective coherent state according to the operators of interest.  The number of $S$ in $\mathscr{S}$ is the number of independent semiclassical quantum fluctuators denoted $N(\epsilon, S)$ in the earlier derivation leading to the estimate 
for $\epsilon$.  This time though, it is independent of $\epsilon$ itself. For example, for a graph that is a cubic lattice having edges parallel to the x,y,z-axes, the surfaces are, respectively, grouped into 3 families each "transverse"  to those axes. In this construction all $I(S)$ contain more than one edge, so all the $S\in \mathscr{S}$ are non-elemental.  For these collective coherent states $\epsilon$ plays the role of a UV cut-off (sliding scale), below which no further geometric structure is resolved by the graph, and not some physically distinct semiclassical or mesoscopic gravitational scale. However, the effective length $\xi$ over which space \emph{fluctuates} (the fluctuation correlation length) is determined by the choice of $\mathscr{S}$ in the state construction, \emph{not} $\epsilon$.  By contrast,  $\epsilon$ corresponds to the typical size of a flat grain of space sampled by a graph, which can be much smaller than $\xi$. What has been realized here is that the original simple tensor product of single edge-based coherent states cannot describe this and is unphysical because the corresponding optimal $\epsilon$ cannot be made sufficiently small to properly
describe  semiclassical matter. Seen from this perspective, collective states, like those OPS have begun to construct, possess a strongly compelling physical necessity, and are not merely an alternative method to construct novel gravitational coherent states.\\
\section{Experimental Consequences Revisited}
How does the absence of the edge length $\epsilon$ as a characteristic physical scale affect the earlier estimates for the space(time) strain spectra? In the equations for the strain spectra $S_{h}(\omega)$, $\epsilon$ is replaced by a scale $\xi$ which is essentially the typical size of the embedded surfaces $S\in\mathscr{S}$ entering the collective coherent state construction. These surfaces now take over the role of the independent quantum fluctuators that the individual edges played previously, so $\xi$ acquires the interpretation of a quantum geometric  correlation length. However, $\xi$ is, so far, not well controlled theoretically, and is known only to lie between $\epsilon$ and $L_{C}$. $\xi$, unlike $\epsilon$, may be larger or smaller than  $\lambda$, the wavelength  of the  probe radiation.  The gradient expansion referred to earlier is vertex based, and so $\xi$ will not be constrained by $\lambda$.  The result for the strain (or 4-metric) noise density is:
\begin{equation}
S_{h}(\lambda , \xi) \simeq (2/\pi)^{1/2} \left(\frac{\delta\xi}{\xi}\right)\left(\Lambda/c\right)^{1/2} \left[ 1+ (2\pi\xi/\lambda)^{2}\right]^{-1/2},
\end{equation}
where $\Lambda$ is the total photon propagation distance, and $\delta\xi$ is the fluctuation in $\xi$.
As previously, it has been assumed that the strain spectrum scales proportionately to $(\delta\xi/\xi)\, (\Lambda/\xi)^{1/2}$.
If we choose to make operators like $\hat E_{j}(S)$ (which has dimensions of area) dimensionless by dividing by $L_{P}^{2}$, then the classicality takes the value $t=8\pi\gamma$, where $\gamma$ is the Barbero-Immirzi parameter.\cite{Flux_Repn_Coh_St} $\gamma$ enters from the Poisson algebra, where $G\gamma$ replaces $G$, the Newton-Cavendish constant. The rms fluctuation of such scaled flux operators has been estimated
 to be $\delta E(S) \simeq t^{1/2}$.\cite{Flux_Repn_Coh_St} On the other hand, for classical gravitational fields of low curvature, $\delta E(S)\simeq 2 \xi \delta\xi/L_{P}^{2}$. This yields the rough estimate 
  \begin{equation}
 \frac{\delta\xi}{\xi} \simeq (1/2) (8\pi\gamma)^{1/2} (L_{P}/\xi)^{2},
   \end{equation}
    and  then
    \begin{equation}
    S_{h}(\lambda ,\xi)\simeq 2\sqrt{\gamma} \left(\frac{L_{P}}{\xi}\right)^{2} \left(\frac{\Lambda}{c}\right)^{1/2} \left[1+(2\pi\xi/\lambda)^{2}\right]^{-1/2}.
    \end{equation}
 If one sets $\gamma= \gamma_{BH}= \ln 2/(\pi\sqrt 3)$, the Bekenstein-Hawking value of the Barbero-Immirzi parameter\cite{Rovelli}, one finds for
 $\xi \lesssim 6.4\cdot 10^{-23}\;\mathrm{m}\; = 4\cdot 10^{12} L_{P}$ (corresponding to energies above $\simeq 2 \cdot10^{4}$ TeV), the 5 GHz Li-Baker apparatus will start to detect these quantum fluctuations, as the quantum gravitational strain fluctuations exceed its (non-propagating) noise floor.\cite{Li_Baker}  Advanced LIGO  using optical photons will begin to sense $\xi \lesssim 5.2\cdot 10^{-25}\;\mathrm{m}\;= 3\cdot10^{10} L_{P}$ (energies above $\simeq 2\cdot10^{6}$ TeV).\cite{LIGO} Both of these lie in the $\lambda$ independent regime of $S_{h}$ and give $\xi$ values sufficiently short  to keep space(time) firmly semiclassical at least beyond the current TeV energy scales of current particle physics experiments. Even if no stochastic space-time strain signals are observed directly by experiments,  this lower bound for $\xi$ can be raised further by  improving the experimental sensitivity bounds. That is, \emph{not }detecting any stochastic space-time strain signal at a given level of strain sensitivity means a \emph{semiclassical} $\xi \gg L_P$ must be \emph{longer} than some length, or that the correlations are not semiclassical at all, i.e. of Planck scale. Of course, detecting a stochastic strain above the experimental noise floor would be even more informative. \\
 
 Here we summarize the key physical assumptions or inputs entering this expression for $S_{h}(\lambda ,\xi)$:
 (1) There exist \emph{semiclassical} fluctuations of 3-geometry which are characterized by a semiclassical (collective) proper length $\xi \gg L_P$, their correlation persistence length as determined from the ambient classical 3-geometry. This means the 3-geometry fluctuations are small in comparison to the ambient 3-geometry, see equations (\ref{SemiclA}, \ref{SemiclB}, \ref{SemiclC}), so we can even speak of a correlation ``length.''  (2) The rms length $\delta\xi$ associated with a single fluctuator of the scale $\xi$ is $\delta\xi\simeq L_P$. (3) The fluctuations
 of 3-geometry are accompanied by fluctuations of 4-geometry with nearly unity projection which are causal in the semiclassical regime. So there is a characteristic high frequency roll-off $\omega _{0}\simeq c/\xi$ in the fluctuation spectrum. (4) The spectrum $S_h(\lambda , \xi)$ of space-time strain or 4-metric fluctuations scales proportionately to $(\delta \xi/\xi)
 (\Lambda/\xi)^{1/2}$, in accord with the Law of Large Numbers, where $\Lambda$ is the distance over which the strain or 4-metric is measured. This expression for the fluctuation spectrum does not depend on how $\xi$ was calculated, that is whether it arises from purely kinematic (near or exact classical expectation values and minimal uncertainty) coherent states or from coherent states also satisfying the quantum constraints, as long as there is indeed a well-defined fluctuation correlation length $\xi\gg L_P$. However, theoretically only kinematic coherent states having such a collective nature have so far been constructed, and they  do not necessarily satisfy those constraints. \\
 
What can be said about an upper bound for $\xi$? As with $\epsilon$, an  upper limit for a semiclassical description of gravity is $\min L_C$, the minimum relevant classical curvature length. So far experimental particle physics has found space-time  to be semiclassical up to the largest laboratory energies attained $E_{PP} \simeq 1$ TeV. A particle having such a rest energy will generate a classical gravitational field at its Compton radius having 
 \begin{equation}
 \min L_C(E_{PP})\simeq \left(\frac{\hbar^{3}c^{7}}{G}\right)^{1/2} \frac{1}{e^{2} E_{PP}(\mathrm{eV})^{2}}\simeq \left(2.41\cdot10^{21}\,\mathrm{m}\right)\,\left(E_{PP}(\mathrm{eV})\right)^{-2}, 
 \end{equation}
 where we have used equation (\ref{LCEQN})  and assumed all the particle's rest mass lies within its Compton radius.  This implies a high energy physics upper bound: 
 \begin{equation}
 \xi \lesssim \min L_{C}(E_{PP}\simeq 1\,\mathrm{TeV}) \simeq 2.41\cdot 10 ^{-3}\,\mathrm{m}, 
 \end{equation}
 which is still macroscopic for  particle physics (''sitting on a Higgs particle''). This upper bound for $\xi$ will shorten in the future as particles of larger rest mass are uncovered. Even though a curvature length in the mm range is short by comparison to astrophysical curvature scales, the gravitational field at the Compton radius is still a weak classical field since the dimensionless classical  gravitational potential there is $\sim 10^{-39}$. If we are simply interested in collective coherent states describing the Earth's gravitational field "sitting in our chairs'' (macroscopic physics), then (\ref{LCEQN}) gives
 $L_C\simeq2.41\cdot10^{11}\, \mathrm{m}\simeq 1.6\,\mathrm{AU}$, literally an astronomically long coherence length.\\
 
However, the current theoretical understanding of collective coherent states is still very rudimentary,  and these simple numerical estimates of  $\xi$ could easily be superseded by other more detailed, precise, and sophisticated approaches.   In particular, it is not clear just how adjustable $\xi$ really is, or even what a reasonable range of values is to assign to it, since it is not yet understood how to choose the surfaces $S\in \mathscr{S}$ that enter the construction of the collective variables that are described by the coherent states. Only future research can resolve these issues. But these crude physical estimates do show that there is a window for $\xi$  such that collective coherent states are at least so far not in conflict with measurements; and a well regulated semiclassical Hamiltonian for classical electromagnetism can also be constructed. \\
 
The estimate for the upper bound on $\xi$ is so long, however, that it practically invites some speculation. Before $\xi$ becomes limited by the minimum classical curvature length, it is possible that another physical mechanism could start to play a role. As with $\epsilon$, the clue may lie in the so far neglected matter sector. Namely,  matter could induce decoherence, dephasing, or decay of purely geometric coherent states. Decoherence and decay phenomena generally occur when one quantum system (A) becomes coupled to to another quantum system (B) that has a large number of degrees of freedom (a ``reservoir'' or ``environment'' for A). This way the quantum information from A that becomes entangled with B's through their mutual coupling gets spread over B's many degrees of freedom, and the probability (amplitude) of A recovering the information leaking to B becomes vanishingly small. This leads to a non-unitary reduced density matrix (decoherence or decay) for A. One well known, experimentally verified example is from atomic physics. Consider an atomic electron in an excited state. Suppose it could make a transition to a lower energy eigenstate by emitting a photon into the electromagnetic radiation field around the atom. The lifetime of the electron in the excited state depends not only on its coupling to the electromagnetic field from its charge, but also on the electromagnetic environment and in particular on the density of radiative states around the energy difference between the excited and lower electron states. If the atom resides inside a electromagnetic cavity where the density of final radiative states is suppressed at that energy difference, or so that there are only evanescent (non-propagating) electromagnetic modes available, then the excited state lifetime is considerably longer than if there are many suitable propagating modes into which photons could be radiated away.  Similarly, electromagnetic vacuum fluctuations lead to a spontaneous emission lifetimes for atomic excited states. One can turn this around, and ask how matter vacuum fluctuations affect coherent sates of the electromagnetic field. Well known results from quantum electrodynamics \cite{Weinberg} have long established that matter loops (radiative corrections) do not give \emph{on-shell} photons a mass, so coherent states of propagating photons do not attenuate through such a mechanism. However, this not the case for \emph{off-shell} photons, whose propagator is affected by radiative corrections. That is, the persistence length of non-propagating quantum electromagnetic fluctuations (not obeying the Maxwell equations) comes from phenomena like vacuum electron-positron pair fluctuations. For the case of quantum gravity, take A to be  a coherent state describing some semiclassical fluctuating collective gravitational degree of freedom. Once matter fields are introduced (system B), quantum geometric information becomes entangled with the matter degrees of freedom. As the number of relevant matter degrees of freedom increases, a similar kind of matter induced decoherence or decay of coherent geometric states could occur. For example,  semiclassical geometric degrees of freedom might become entangled with a thermal gas of photons or with particles/fields whose decay products are distributed over a large phase space. Additionally, vacuum fluctuations of matter degrees of freedom could act on off-shell geometry fluctuations.  That is, matter fields could lower the upper bound for $\xi$ as the persistence length of quantum gravitational \emph{fluctuations} becomes limited before the general relativistic curvature bound $\xi\lesssim \min L_{C}$ is attained.  Put differently, matter might play a significant role in determining how long semiclassical purely gravitational fluctuations (not obeying Einstein's field equations) can cohere.  Unfortunately the understanding of matter and its quantum fluctuations in LQG is presently too undeveloped to start quantifying such physical ideas, and methods from standard quantum field theory are obviously not directly applicable.\\ 
 
As mentioned in the introduction, another area of potential contact between LQG theory and measurement is the  observational search for the predicted photon vacuum dispersion. One way to derive this effect is to utilize tensor product coherent states to calculate
expectation values of gravitational degrees of freedom, and the edge length's dependence on $L_{P}$ then enters the final coefficients of the photon dispersion relation $\omega(k)$. In Sahlmann's discussion of this phenomenon\cite{Sahl_Thesis}, the $\epsilon$ optimization procedure
sketched earlier motivates parametrizing $\epsilon$ as $\epsilon\simeq L_{P}^{\beta} L_{C}^{1-\beta}$ for some fixed exponent $\beta \in (0,1)$. STW found $\beta =1/3$, and consequently predicted that the coefficients
in the photon vacuum dispersion should have a non-integer power dependence on $L{_P}$.  However, now that it has been realized that the optimization procedure leads to non-physical values for the edge length relative to photon wavelengths, that prediction is drawn into question. The gravitational expectation values will have to be calculated using collective coherent states, where $\epsilon$ plays a different role.
Again, one expects the length $\xi$ to be the dominant actor, and that length is not initially anticipated to have any  fractional power dependence on $L_{P}$. If that is true, the application of  collective coherent states to the problem would suggest
that the photon vacuum dispersion coefficients should display an integer power dependence on $L_{P}$, similar to the earlier results of
\cite{GP} and \cite{ATU}. However, a definitive conclusion must await explicit calculations of the photon dispersion using collective coherent states as well as the $L_{P}$ dependence of $\xi$. \\
\section{Conclusions}
We have studied the physical consequences of an optimized edge length $\epsilon$ in LQG coherent states. It was found that the tensor product over edges of the graph forces $\epsilon$ to 
lie in the range $100\,\mu\mathrm{m} - 0.7\,\mathrm{m}$ on Earth.  While the theoretically implied space(time) strain fluctuations about these states are still experimentally out of reach, examining the semiclassical Maxwell Hamiltonian reveals difficulties in having both such a long edge length and a self-consistent regularization scheme for classical electromagnetism. Instead, the edge length does not determine a characteristic physical scale like the coherence or magnetic penetration lengths from superconductivity. Rather, it is a sliding/running scale that serves as a cut-off governing how much detailed geometric information about the continuum geometry becomes encoded into an embedded graph,  so then a self-consistent regulated semiclassical matter Hamiltonian can be constructed. The non-edgewise tensor product, kinematic collective coherent states recently developed by Oriti, Pereira, and Sindoni \cite{OPS} capture this essential physics, and the typical size of the embedded 2-surfaces entering that collective description then delineate a characteristic mesoscopic scale $\xi$ for semiclassical LQG. However, these states do not necessarily satisfy the quantum constraints. Put succinctly, $\epsilon\gtrsim L_{P}$ is the adjustable size of a flat chunk of space sensed by graphs, which should be taken smaller than any relevant matter probe wavelength. In this sense $\epsilon$ is similar to the sliding energy scale $\mu$ familiar from renormalization group methods in quantum field theory. On the other hand,  $\xi > \epsilon$ is the correlation length for quantum fluctuations of semiclassical geometry. Matter interactions (measurement of geometry by matter fields) could play a role in determining the collection of 2-surfaces $\mathscr{S}$ entering the construction of the semiclassical gravitational collective coherent states, and thereby affect $\xi$ as well. As pointed out in \cite{Th_CompStates}, matter can only reside where geometry gives it a place to live, so surfaces or loops defined by matter will lie within the graph and could provide a natural set of surface/loop-based operators with respect to which the set of coherent states are to behave 
semiclassically. However many open theoretical questions remain to be answered before detailed predictions can be made. \\

In spite of the embryonic state of theoretical affairs here, the phenomenology suggests the possibility of at least placing tighter experimental lower bounds on $\xi$ and directly to learn more about semiclassical space-time. Experimental
efforts probing space-time strain and 4-geometry fluctuations, and in particular those utilizing the Li-Baker method, are encouraged to analyze (or place bounds on) any stochastic signals (noise of non-instrumental, non-artifact origin) from this perspective. We won't learn more unless we look.\\

\section{Acknowledgements}
The author wishes to thank Prof. Zhen Yu Zhang of USTC, Hefei, China for encouraging him to revisit this problem, S.T. Lu for stimulating discussions, and P.D. Lu for his kind assistance in Shanghai where some of this work was carried out.\\

\end{document}